\documentclass[journal=nalefd,manuscript=letter]{achemso}

\usepackage[version=3]{mhchem} 
\usepackage{lmodern}

\usepackage[utf8]{inputenc}

\usepackage{textcomp}

\newcommand*\saguyref[2]{\hbox{-S}}

\author{Felipe Fávaro de Oliveira}%
 \affiliation{$3.$ Institute of Physics, Research Center SCoPE and IQST, University of Stuttgart, 70569 Stuttgart, Germany}
 \email{f.favaro@physik.uni-stuttgart.de}
\author{S. Ali Momenzadeh}%
 \affiliation{$3.$ Institute of Physics, Research Center SCoPE and IQST, University of Stuttgart, 70569 Stuttgart, Germany}
\author{Denis Antonov}%
 \affiliation{$3.$ Institute of Physics, Research Center SCoPE and IQST, University of Stuttgart, 70569 Stuttgart, Germany}
\author{Jochen Scharpf}
 \affiliation{Institute for Quantum Optics, University of Ulm, 89081 Ulm, Germany}
\author{Christian Osterkamp}
 \affiliation{Institute for Quantum Optics, University of Ulm, 89081 Ulm, Germany}
\author{Boris Naydenov}
 \affiliation{Institute for Quantum Optics, University of Ulm, 89081 Ulm, Germany}
\author{Fedor Jelezko}
 \affiliation{Institute for Quantum Optics, University of Ulm, 89081 Ulm, Germany}
\author{Andrej Denisenko}%
 \email{a.denisenko@physik.uni-stuttgart.de}
 \affiliation{$3.$ Institute of Physics, Research Center SCoPE and IQST, University of Stuttgart, 70569 Stuttgart, Germany}
\author{Jörg Wrachtrup}%
 \affiliation{$3.$ Institute of Physics, Research Center SCoPE and IQST, University of Stuttgart, 70569 Stuttgart, Germany}
 \alsoaffiliation{Max Planck Institute for Solid State Research, 70569 Stuttgart, Germany}

\title{Towards optimized surface $\delta$-profiles of nitrogen-vacancy centers activated by helium irradiation in diamond}

\abbreviations{nitrogen-vacancy center (NV center), nuclear magnetic resonance (NMR), magnetic resonance force microscopy (MRFM), spin coherence time (T${}_2$), chemical vapor deposition (CVD), inductively coupled plasma (ICP), spin-lattice relaxation time (T${}_1$), }

\begin{document}

\begin{tocentry}
\includegraphics{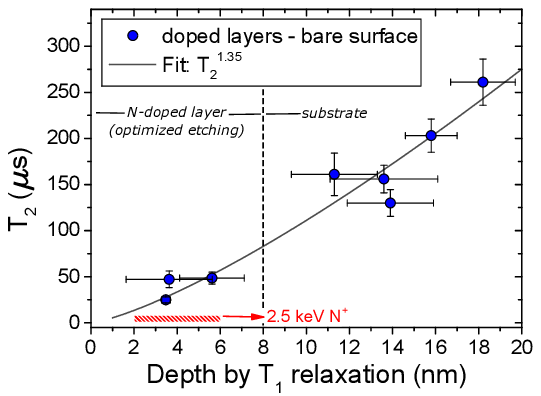}
\end{tocentry}

\begin{abstract}

The negatively-charged nitrogen-vacancy (NV) center in diamond has been shown recently as an excellent sensor for external spins. Nevertheless, their optimum engineering in the near-surface region still requires quantitative knowledge in regard to their activation by vacancy capture during thermal annealing. To this aim, we report on the depth profiles of near-surface helium-induced NV centers (and related helium defects) by step-etching with nanometer resolution. This provides insights into the efficiency of vacancy diffusion and recombination paths concurrent to the formation of NV centers. It was found that the range of efficient formation of NV centers is limited only to approximately $10$ to $15\,$nm (radius) around the initial ion track of irradiating helium atoms. Using this information we demonstrate the fabrication of nanometric-thin ($\delta$) profiles of NV centers for sensing external spins at the diamond surface based on a three-step approach, which comprises (i) nitrogen-doped epitaxial CVD diamond overgrowth, (ii) activation of NV centers by low-energy helium irradiation and thermal annealing, and (iii) controlled layer thinning by low-damage plasma etching. Spin coherence times (Hahn echo) ranging up to $50\,$\textmu s are demonstrated at depths of less than $5\,$nm in material with $1.1\,\%$ of $^{13}$C (depth estimated by spin relaxation (T$_1$) measurements). At the end, the limits of the helium irradiation technique at high ion fluences are also experimentally investigated.

\end{abstract}


Sensing weak magnetic fields originating from single or small ensembles of targeted spins has been a subject of intense research over the past years. In particular, the negatively-charged nitrogen-vacancy (NV) center hosted in diamond\cite{reviewJWrachtrup,ReviewDoherty} has been proposed\cite{CaiTheorSMol,PerinicicTheorNMR,KostTheorSMole} and successfully employed as a magnetic field sensor for nuclear\cite{MaminNMR2013,staudacher,NMRMueller, Lorentz1.9nm,haeberle} and electronic\cite{Bernhard,MaminDEER,singleProteinFazhan} external spins. Most important, room temperature operation featuring nanometric resolution and sensitivities below $\frac{nT}{\sqrt{Hz}}$ have been shown\cite{TaylorSens,haeberle,GopanBsens,TWolfMagnetometry}, meaning major advances in comparison to other techniques such as standard nuclear magnetic resonance (NMR)\cite{TyszkaNMRfirst,GloverNMR} and magnetic resonance force microscopy (MRFM)\cite{SidlesMRFMfirst,DegenMRFM}. In most applications the coupling between the NV center (sensor) and targeted external spins relies on their weak magnetic dipolar interaction. Consequently, NV centers must be located in nanometer vicinity of the diamond surface, as the coupling strength decays with the third power of the distance (r$^3$) between the sensor and the targeted spins\cite{MaminNMR2013,staudacher,NMRMueller, Lorentz1.9nm}. Furthermore, the related detection sensitivity scales proportionally to $\sqrt{T{}_2}$\cite{Maze_sqrtT2}, implicating that long spin coherence times (${T}_2$) are highly desired.

So far, generation of near-surface NV centers has relied primarily on low-energy nitrogen implantation followed by thermal annealing\cite{RabeauLabN15,Ofori-OkaiSPVeryShallow,pezzagnaYield}. Limitations of this technique are mainly low yield\cite{pezzagnaYield} and strongly degraded spin properties\cite{Ofori-OkaiSPVeryShallow,staudacher,Lorentz1.9nm} in comparison to single NV centers in bulk diamond. In addition, nanoprecise control in the depth of created NV centers remains a challenge as a result of ion straggle in the z direction. To reduce the decoherence effect from implantation-induced defects, NV centers can be created alternatively by irradiating nitrogen-doped diamond layers with electrons, protons or inert ions to create vacancies. This method relies on thermal diffusion of vacancies from the initial tracks of the irradiating species, thus leading to the formation of NV centers in a relatively defect-free environment. By using such combined methods, the creation of NV centers with improved optical\cite{AcostaEIRR,AharonovichIR,FCWHeIrr,HuangHeIrrMic} and spin\cite{dGrownT2_KOhno,KOhnoC12_3D} properties have already been demonstrated. Nevertheless, optimized conditions for the engineering of a targeted structure, namely a nanometric-thin surface layer containing NV centers suitable for sensing applications, have not been demonstrated yet.

Among other constrains, activation of NV centers beyond the nitrogen-doped targeted surface layer is unavoidable when its thickness is much smaller than the projected depth range of irradiation-induced vacancies\cite{dGrownT2_KOhno}. Alternative methods have been proposed\cite{KOhnoC12_3D}, but quantitative knowledge about the activation of NV centers by ion irradiation and thermal annealing still restrains further optimization of this technique. Investigations reported so far indicate that the conversion efficiency from ingrown nitrogen atoms to NV centers by such a method decreases rapidly with increasing separation between the vacancy sources and targeted nitrogen-doped layer. For instance, in a structure where vacancies are created in a buffer layer at approximately $50\,$nm away from the nitrogen-doped region, the conversion efficiency was shown to be about $1\,\%$\cite{KOhnoC12_3D}. In contrast, vacancies created directly within nitrogen-doped layers have demonstrated an efficiency of approximately $10\,\%$ with proper charge stability\cite{HuangHeIrrMic}, although NV centers located in the close vicinity of implantation-induced defects are expected to have degraded spin properties. It is therefore of great interest to have insights into the vacancy diffusion process in diamond associated to the creation of NV centers at the nanoscale, especially in the near-surface region. In this report we address this issue, presenting a detailed analysis of the depth profiles of NV centers created by low-energy helium irradiation and thermal annealing in nitrogen-doped chemical vapor deposition (CVD) diamond layers. The details derived from the profiles of helium-induced NV centers were applied to optimize a novel approach to engineer nanometric-thin ($\delta$) profiles of NV centers embedded directly at the diamond surface. The experimental realization of such a $\delta$-structure and related limits of the ion irradiation technique will be discussed in this letter.


\begin{figure}
\includegraphics{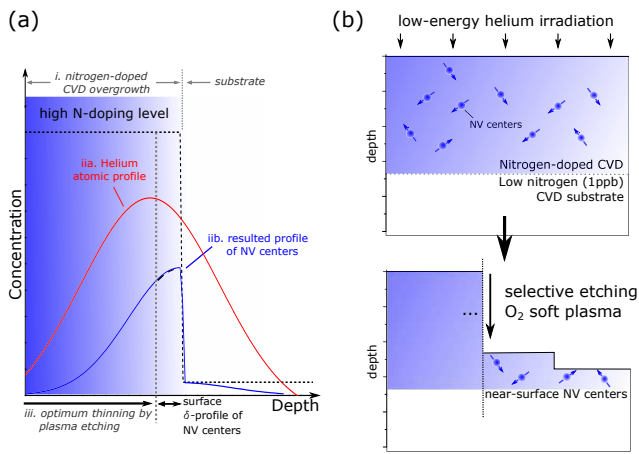}
\caption{\label{figure1} (a) Schematic representation of the proposed method for the engineering of $\delta$-profiles of NV centers at the diamond surface. The final etching step brings the maximum of the profile of NV centers to the diamond surface for external spins sensing. (b) Corresponding experimental method.}
\end{figure}

To explore further the advantages of combining ion irradiation and nitrogen-doped diamond layers, a new approach is proposed. Figure \ref{figure1} shows a schematic representation of such a structure and the related fabrication steps. At first, a thin diamond film with controlled nitrogen doping is grown on a diamond substrate with low content of nitrogen impurities. As a consequence, a sharp profile of nitrogen doping is created at the interface between the substrate and overgrown layer, leading eventually to a sharp cutoff in the corresponding profile of NV centers (figure \ref{figure1}(a)). As the second step, a profile of vacancies is created by low-energy helium irradiation, activating NV centers from ingrown nitrogen atoms during the subsequent thermal annealing. An important issue regarding the fabrication process of the proposed structure is the initial thickness of the overgrown layer. Indeed, this parameter should correspond to the peak in the depth profile of created NV centers in order to have an optimum density of NV centers at the final surface (see figure \ref{figure1}(a)). The location of this peak is given not only by the helium atomic profile, but also by the diffusion length of vacancies during thermal annealing, a feature that has not been yet quantitatively explored at the nanoscale. As the third and final step, low-damage, nanometric-precise oxygen plasma is applied to remove a part of the overgrown layer, thus bringing the maximum density of NV centers directly to the surface for efficient detection of external spins. In this way, ion straggle in the z direction is taken into account and the corresponding depth uncertainty can be minimized. The applied etching process is based on an oxygen inductively coupled plasma (ICP), whereby nanometric-precise etching was shown while preserving the optical and spin properties of near-surface NV centers\cite{OxygenSoftPlasma}.

As already mentioned, the diffusion of vacancies in diamond in regard to the efficient formation of NV centers by ion irradiation is yet to be explored. Therefore, before introducing the experimental realization of such a $\delta$-structure, aspects regarding the formation of NV centers by low-energy ion irradiation such as depth profiles and conversion efficiency were investigated using a reference sample, namely a CVD diamond substrate with a low and homogeneous content of nitrogen impurities (\textless $1\,$ppb). Individual areas of the sample were irradiated with helium ions (He$_2^+$, $4.0\,$keV - fluence of $10^{11}\,$cm${}^{-2}$), which resulted in single NV centers after thermal annealing. In this fluence range, the density of NV centers was shown to increase linearly, as will be further discussed in this letter. Afterward, each irradiated region was individually etched in plasma to a specific depth and the related areal densities of NV centers were measured by confocal microscopy technique. Further details in regard to sample preparation, characterization and related features can be found in the Supporting Information (SI).

\begin{figure}
\includegraphics{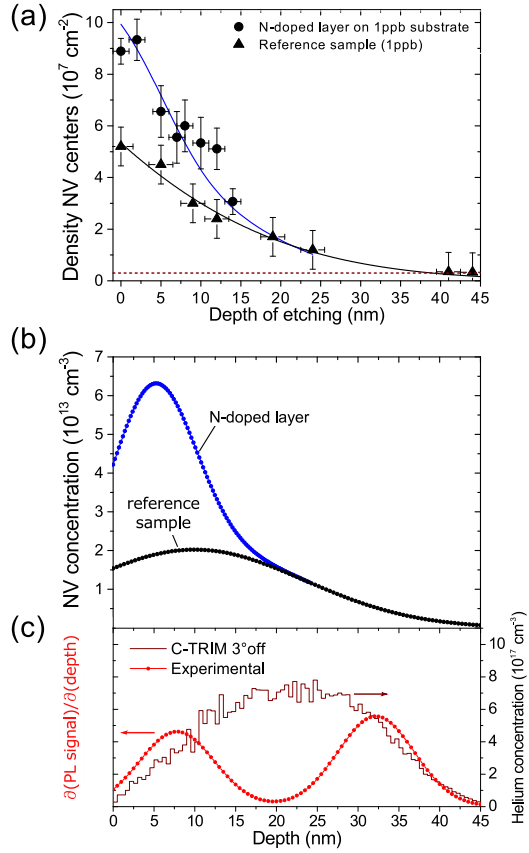}
\caption{\label{figure2} (a) Areal density of NV centers vs. the etching depth for two samples: substrate with low and homogeneous content of nitrogen (triangles) and an overgrown nitrogen-doped layer (circles); the solid lines are fits to Gaussian complementary error functions; helium ion fluence is 10${}^{11}\,$cm${}^{-2}$ and the dashed line represents the sample background (non-irradiated regions). (b) Related experimentally fitted depth profiles of NV centers; (c) atomic profile simulated by C-TRIM and experimental profile of helium-related defects at the ion fluence of $2 \times 10^{12}\,$cm${}^{-2}$.}
\end{figure}

As described, the areal density of NV centers through the etching steps is shown in figure \ref{figure2}(a), represented by the black triangles. The experimental values were fitted for simplicity to a Gaussian complementary error function, whereby the mean ($\mu$) and standard deviation ($\sigma$) fitted values determine the related Gaussian distribution function\cite{abramowitz_stegun,OxygenSoftPlasma}. Such an experimentally fitted depth profile of NV centers is shown in figure \ref{figure2}(b) (reference sample). Furthermore, helium irradiation with ion fluences higher than 10${}^{12}\,$cm${}^{-2}$ performed on other regions of the sample has led to the appearance of a helium-related photoluminescence (PL) background in confocal microscopy measurements (see SI). This feature allowed the evaluation of the depth profile of irradiating helium atoms (through helium-related defects) by plasma etching simultaneously, as presented in figure \ref{figure2}(c). For comparison, the related helium atomic profile simulated by Crystal-TRIM (CTRIM)\cite{CTRIM} is also shown. Here, the ion channeling effect\cite{WilsonChanneling,ToyliChanneling} was taken into account for a [100]-oriented diamond crystal with a 3$^{\circ}$-off implantation angle. As can be seen, the channeling tail of the profile of irradiating atoms should also be taken into account considering the depth distribution of vacancies for the formation of NV centers. In addition, a similar observation has been reported for diamonds implanted with low-energy nitrogen ions\cite{OxygenSoftPlasma}. 

Based on the described results, the conversion efficiency from ingrown nitrogen atoms to NV centers is estimated to be $15\,\pm 5\,\%$ for nitrogen doping in the low ppb range, which is in good agreement with the literature\cite{HuangHeIrrMic} and represents a noticeable improvement in comparison to standard nitrogen implantation at comparable depth ranges\cite{pezzagnaYield}. The uncertainty in this estimation can be related directly to the estimated value of the nitrogen content in the diamond substrate (see SI). Furthermore, comparing the depth profiles of irradiating helium atoms and helium-induced NV centers in figure \ref{figure2}(b) and (c) shows that NV centers are created efficiently only within the range of the helium atomic profile. In fact, the creation of most NV centers by helium irradiation is restrained to distances of approximately $10\,$nm (radius) from the initial (as-implanted) ion tracks. This is in contrast to what has been usually considered, where the range related to the formation of NV centers by irradiation techniques is believed to be comparable to the diffusion length of vacancies in diamond, which might be extended to several hundreds of nanometers under the used annealing conditions\cite{KOhnoC12_3D,OrwaEa}. This contradiction can be explained considering each individual ion tracks to behave as point-like sources of vacancies. During thermal annealing, the radial distribution of vacancies $Vac(r)$ in a quasi-3D space should yield a \textit{strong reduction} in the related concentration within the first nanometers of increasing distance $r$ from ion tracks\cite{Chatterjee2002201}. On the other hand, the volumetric probability $P(r)$ to find an ingrown nitrogen atom in a volume surrounding the initial ion track increases with increasing $r$. As a consequence, the probability to create a single NV center within such a volume should be proportional to the product $Vac(r) \times P(r)$. In this case, the probability of creation of NV centers would be limited to a nanometric volume around the initial ion tracks, as indeed observed in the presented experiment (figure \ref{figure2}), thus defining an \textit{effective volume of creation of NV centers}. However, a precise estimation of the corresponding diffusion and recombination of vacancies in a quasi-3D space (at the proximity of the diamond surface) is above the scope of this letter.

The obtained results have provided so far crucial information towards optimized irradiation and growth conditions for the proposed $\delta$-structure in figure \ref{figure1}(a). In the second experiment, a nitrogen-doped CVD layer was grown by microwave-assisted CVD plasma on the surface of an identical diamond substrate containing low concentration of nitrogen impurities (\textless $1\,$ppb). Importantly, the thickness of the overgrown nitrogen-doped layer was set to be close to the maximum of the profile of NV centers presented in figure \ref{figure2}(b). Further details of the CVD overgrowth process can be found in the SI. The areal density of NV centers as a function of depth is shown in figure \ref{figure2}(a). The corresponding fit is a Gaussian complementary error function in addition to the reference sample, as described above. The uncertainty in the experimental data can be related to non-homogeneous incorporation of nitrogen during overgrowth in different areas of the sample. Further, the corresponding depth profile of NV centers is shown in figure \ref{figure2}(b). A peak in the concentration of NV centers identifies the high level of nitrogen doping in such a layer. Thereby, the concentration of nitrogen impurities in the overgrown layer was estimated to be about $10\,$ppb with thickness in the range of $15$ to $20\,$nm. Eventually, approximately $10\,$nm were etched from a part of the original layer in order to optimize the structure and thus bring the calibrated peak of NV centers directly to the surface. The spin properties of these NV centers and their corresponding depth are described in the following section.

\begin{figure}
\includegraphics{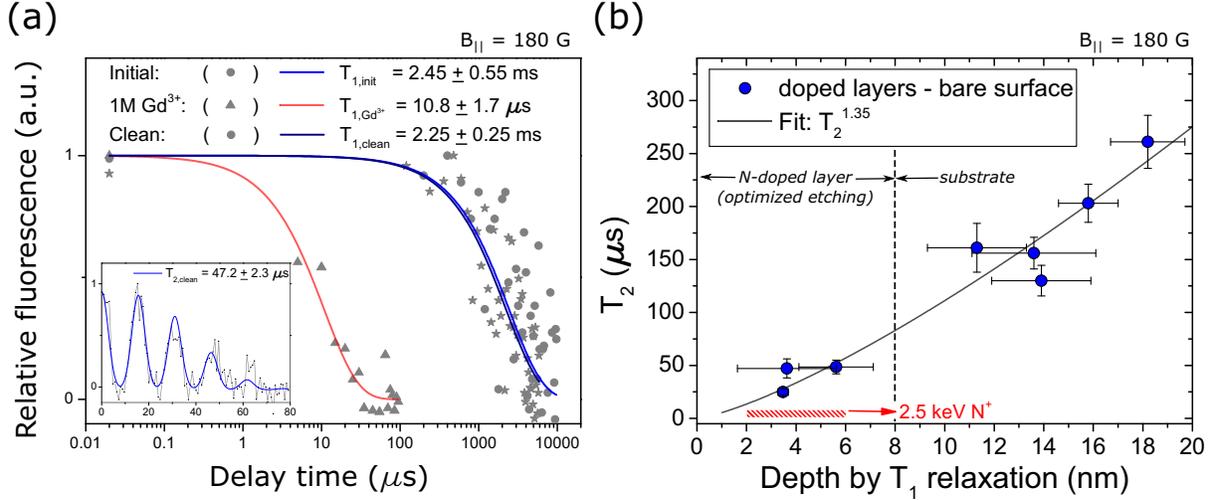}
\caption{\label{figure3} Evaluation of the $\delta$-layer after optimized etching, as proposed in figure \ref{figure1}. (a) T${}_1$ time of an individual NV center within the optimized surface layer; the solid curves fit the data points in the presence and absence (initial and clean) of the Gd${}^{3+}$ layer. The inset shows the corresponding T${}_2$ time. (b) T${}_2$ times vs. depth of individual NV centers; depth values were obtained by sensing Gd${}^{3+}$ ions at the diamond surface. The highlighted region in red represents roughly the typical range for the case of NV centers created by nitrogen implantation with $2.5\,$keV of energy.}
\end{figure}

After the described plasma etching step, the remaining NV centers were expected to be distributed within a layer of $5$ to $8\,$nm in thickness, according to the data presented in figure \ref{figure2}(b). It is known that magnetic noise originated from spins on the diamond surface can be used to precisely define the depth of individual NV centers (see SI)\cite{staudacher,NMRMueller,SteinertGd}. In the presented experiment, Gd${}^{3+}$ ions were embedded in a thick (approx. $1\,$\textmu m) layer on top of the diamond surface. Such technique is based on the magnetic noise from external electronic spins, which gives a higher dynamic range than NMR-based methods. Therefore, spin-lattice relaxation times (T${}_1$) from individual NV centers were measured in the absence and presence of such a layer, as presented in figure \ref{figure3}(a). Thereby, Gd${}^{3+}$-induced spin relaxation is extracted and used to estimate the corresponding depth of NV centers as described by S. Steinert \textit{et al.}\cite{SteinertGd}. During this evaluation procedure, corresponding spin coherence times (T${}_2$) were also measured and are summarized in figure \ref{figure3}(b). The results show a decrease in T${}_2$ times for lower depths, which can be attributed to stronger interactions with spins at the diamond surface\cite{PI3overgrowth,MyersNoiseCalibNVs}. The experimental fit (depth $\sim$ T${}_2^{1.35}$) is evaluated regarding the model for the depth evaluation, as described in the SI. The NV centers located within the optimized surface layer at depths of less than $5\,$nm have shown T${}_2$ times of up to $50\,$\textmu s, corresponding to a fivefold improvement in comparison to NV centers created by nitrogen implantation with $2.5\,$keV of implantation energy\cite{PI3overgrowth} (red marked region in figure \ref{figure3}(b)). It must be highlighted that the CVD overgrown layer contains natural abundance of $^{13}$C atoms ($1.1 \%$). In addition, the comparison with T${}_2$ times measured in the substrate (figure \ref{figure3}(b)) assures the crystal quality of the overgrown layer. As a projection for the presented case, it is known that the minimum detectable ac magnetic field by near-surface NV centers is given by $B_{min}$ $\sim (1.7837\times 10^{-10})(1/\sqrt{T T_2})$, where T represents the acquisition time\cite{TaylorSens}. Assuming T${}_2 = 50\,$\textmu s and a typical value of T $= 60\,$s, the minimum detectable ac magnetic field is estimated to be approximately $3.3\,$nT, being sufficient to detect the signal of a single proton spin on the diamond surface\cite{OhashiC12NMR}.

\begin{figure}
\includegraphics{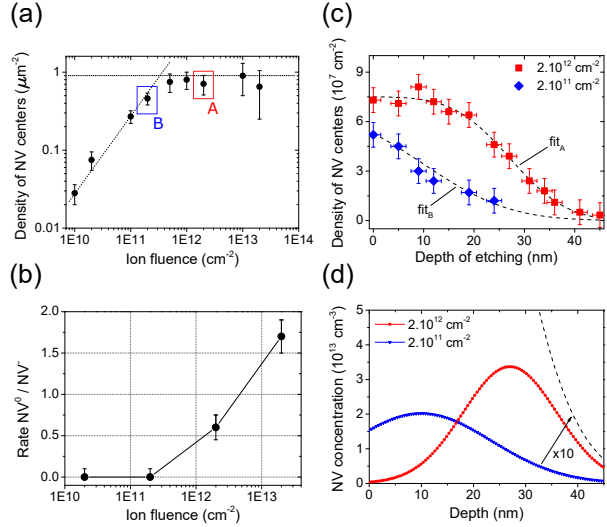}
\caption{\label{figure4} Limits of helium irradiation technique: (a) areal density of NV centers induced by different ion fluences. (b) Ratio NV$^0$/NV$^-$ for different ion fluences. (c) Areal density of NV centers vs. depth of etching and (d) corresponding depth profiles of NV centers for two different ion fluences (see main text for details).}
\end{figure}

The limits of helium irradiation technique were explored in this study by analyzing the ion fluence dependency of the number of helium-induced NV centers combined with depth profiling by means of plasma etching. The investigation was done using the described CVD diamond with low and homogeneous content of nitrogen impurities. Figure \ref{figure4}(a) presents the experimental areal density of NV centers induced by different ion fluences after thermal annealing. Although the number of created vacancies is expected to increase monotonically with the number of irradiating atoms, the experimental density of NV centers shows a saturation effect, observed for ion fluences higher than approximately $5 \times 10^{11}\,$cm${}^{-2}$. Simulations using the stopping and range of ions in matter (SRIM)\cite{SRIM2010} results in vacancy concentration values of 10${}^{18}$ - 10${}^{20}\,$cm${}^{-3}$ within the used ion fluence range. This is approximately two orders of magnitude lower than the damage threshold for diamond amorphization\cite{KalishDamThresh,Uzan-SaguyDamageT} and therefore cannot explain this saturation behavior. A similar effect has been reported for helium irradiation with $30\,$keV of energy\cite{HuangHeIrrMic}, meaning that this is not induced by vacancy losses through surface recombination. At the same time, the contribution of the neutral (NV$^0$) PL signal in the PL spectrum of individual NV centers is shown to increase with the ion fluence, as illustrated in figure \ref{figure4}(b).

To get more insights into this saturation effect, depth profiling (method as described for figure \ref{figure2}) was performed on two regions irradiated with $2 \times 10^{11}$ and $2 \times 10^{12}\,$cm${}^{-2}$ ion fluences, representing linear and saturated regimes in respect to figure \ref{figure4}(a), respectively. These results are summarized in figure \ref{figure4}(c) and (d). In the latter, the dashed line represents the profile expected for the $2 \times 10^{12}\,$cm${}^{-2}$ ion fluence. However, the experimental profile shows a strong suppression in the formation of NV centers in a sub-surface region of less than $25\,$nm in depth. 

The sub-surface suppression and the corresponding saturation effect can be related to activation of additional paths of vacancy recombination during thermal annealing: according to SRIM simulations, the near-surface region hosts most of the created vacancies by ion impact. At depths of $20-30\,$nm, the onset of the suppression effect in figure \ref{figure4}(d), the average distance of irradiating helium atoms is in the range of $10\,$nm for the given ion fluence ($2 \times 10^{12}\,$cm${}^{-2}$). This distance is actually less than the expected lateral struggling for irradiation of molecular helium ions with $4.0\,$keV of energy (SRIM; approximately $15\,$nm). Since the radius of the corresponding effective volume of creation of NV centers was estimated to be approximately $10\,$nm, such overlap of neighbor ion tracks should increase the probability of formation of other vacancy-related defects such as di-vacancies (V$_2$) in the sub-surface region\cite{Annealing1000}. This, in turn, will lead to a reduced number of vacancies available for the formation of NV centers. Moreover, such vacancy-related defects are known to create deep levels in the band gap of diamond, resulting in the depletion of the charge state of NV centers\cite{DeakV2Calc}. This may explain the increased contribution of NV$^0$ during optical excitation in the presented experiment (figure \ref{figure4}(b)). Furthermore, the increasing helium-related fluorescence with ion fluence reinforces the assumption of activation of alternative vacancy recombination paths (see SI).

In conclusion, a detailed analysis of the activation of NV centers by low-energy helium irradiation was presented, revealing information regarding the vacancy diffusion and recombination processes in diamond at the nanoscale. Based on that, a novel approach to engineer $\delta$-profiles of NV centers at the surface was proposed and experimentally verified. Single NV centers were observed with a fivefold improvement in the spin coherence times even in the nanometer vicinity of the etched surface, demonstrating potential capabilities for detection with single-spin sensitivity. In particular, augmented sensitivity would be possible by integrating such technique to photonic nanostructures, such as tapered nanopillars\cite{AliPillar}, and also CVD overgrowth with ${}^{12}$C isotopically purified material.

\begin{acknowledgement}


The authors thank the funding support from the EU/SQUTEC, SIQS, DFG Forschergruppe 1493, MPG and Baden Württemberg Stiftung. F. Fávaro de Oliveira acknowledges CNPq for the financial support through the project No. 204246/2013-0.

\end{acknowledgement}

\begin{suppinfo}
\setcounter{figure}{0}
\renewcommand{\thefigure}{SI-\arabic{figure}}

(1.) Methods, (2.) samples, (3.) sample characterization, (4.) limits of helium irradiation and (5.) Sensing Gd${}^{3+}$ by longitudinal spin relaxation.

\section{Methods}

In the presented experiments, irradiation of He\({}_2^+\) molecular ions with $4.0\,$keV of energy within the described ion fluences was performed. Subsequently, samples were submitted to thermal annealing at $950\,{}^{\circ}$C in high vacuum (below $10^{-6}\,$mbar) for two hours. The annealing parameters were chosen to maximize the yield of negatively-charged NV centers\cite{OrwaATemp} and reduce the concentration of irradiation-induced paramagnetic defects\cite{Annealing1000}. Further, boiling in a triacid mixture ($H_2SO_4:HNO_3:HClO_4,$ $1:1:1$ volume ratio) was used to clean and oxygen-terminate the diamond surface.

\section{Samples}

\subsection{Diamonds with low content of nitrogen impurities}

Samples referred in the main text to have low content of nitrogen impurities were $[100]$-oriented electronic grade ultra-pure (\textless $5\,$ppb content of nitrogen, specified) single-crystal diamonds with natural abundance ($1.1\%$) of ${}^{13}$C commercially available from Element Six. The initial as-polished surface had a root mean square (RMS) roughness of approximately 1 nm, measured by atomic force microscopy (AFM).

A critical aspect related to the conversion efficiency by ion irradiation reported in this letter is an accurate estimation of the initial concentration of nitrogen impurities in this type of sample. It is known that during the CVD growth nitrogen atoms are converted to ingrown NV centers in a rate of approximately $300:1$\cite{E6NtoNVratio}. Thus, confocal microscopy measurements were performed at different depths (up to $60\,$\textmu m) to extract the average concentration of ingrown NV centers, yielding a nitrogen concentration of $0.8 \pm 0.2\,$ppb. This value is in good agreement with the measurements reported by T. Yamamoto et al.\cite{Annealing1000} on identical diamond materials. The reported conversion efficiency of $15 \pm 5\,\%$ was obtained by correlating this estimated nitrogen content with the maximum concentration of NV centers by helium irradiation with $2 \times 10{}^{12}\,$cm${}^{-2}$ ion fluence (figure \ref{figure4}(d) - main text).

\subsection{Nitrogen-doped CVD diamond overgrown layer}

For the structure investigated in this letter (figure \ref{figure1} in the main text), a CVD diamond with low content of nitrogen and fine-polished surface was taken as substrate (see above). Overgrowth of the nitrogen-doped layer was performed by microwave-assisted chemical vapor deposition\cite{PI3overgrowth}. The growth conditions were the following: $1.2\,$kW microwave power, $750\,{}^{\circ}$C growth temperature, $26\,$mbar chamber pressure, $300\,$sccm H${}_2$ flow, $0.4\,$sccm CH${}_4$ (natural abundance of ${}^{13}$C, $1.1\,\%$) flow, $40\,$sccm N${}_2$ flow and 4 minutes of growth time. The expected growth rate was calibrated on other samples with nitrogen-free overgrown CVD layers by weight difference before and after the growth process. Due to the large chamber volume of the CVD reactor, the saturation conditions for the species in plasma were not reached during the used growth time. The concentration of nitrogen impurities with the introduction of nitrogen-doping gas remains therefore not fully calibrated.

For the evaluation of the depth profiles described in the main text, a sequence of plasma step-etching was performed in different areas of the sample, whereby the density of NV centers was traced in individual areas through the etching steps. It must be highlighted that the chosen plasma process allows high selectivity by means of masking the diamond surface with lithographically-patterned photoresist, permitting a large number of experiments to be realized in one sample.

At last, many NV centers have shown blinking behavior, indicating charge instability due to the presence of Gd${}^{3+}$ ions during the measurements of spin relaxation, as described in the main text. Therefore, the sample surface was terminated with fluorine using the plasma process described by C. Osterkamp \textit{et al.}\cite{OsterkampFluor}, after which no blinking could be observed. Importantly, T${}_1$ and T${}_2$ times have shown no remarkable changes with oxygen or fluorine terminations for the bare surface .

\section{Sample characterization}

The samples were characterized using a home-build confocal microscopy setup under ambient conditions. A $532\,$nm wavelength laser was used for excitation and the NV centers emission was filtered by a $650\,$nm long pass filter. In order to identify single emitters, the photoluminescence (PL) signal was detected by two avalanche photo-diodes for second-order autocorrelation $g^{(2)}(\tau)$ measurements\cite{Kurtsiefer_g2}. Additionally, PL spectra were acquired using a SP2300 spectrograph (Princeton Instruments) coupled to a multi-mode fiber, where a $575\,$nm long pass filter was used to filter the excitation laser. 

For coherent manipulation, microwave pulses were generated by programmed FPGAs and delivered on the sample by a $20\,$\textmu m copper wire located on the surface vicinity of the targeted region. A magnetic field of approximately $180\,$Gauss was applied aligned to the NV center axis\cite{DreauBfieldal} for measurements of T${}_1$ and T${}_2$ (Hahn echo).

\section{Limits of helium irradiation: helium-related defects}

\begin{figure}
\includegraphics{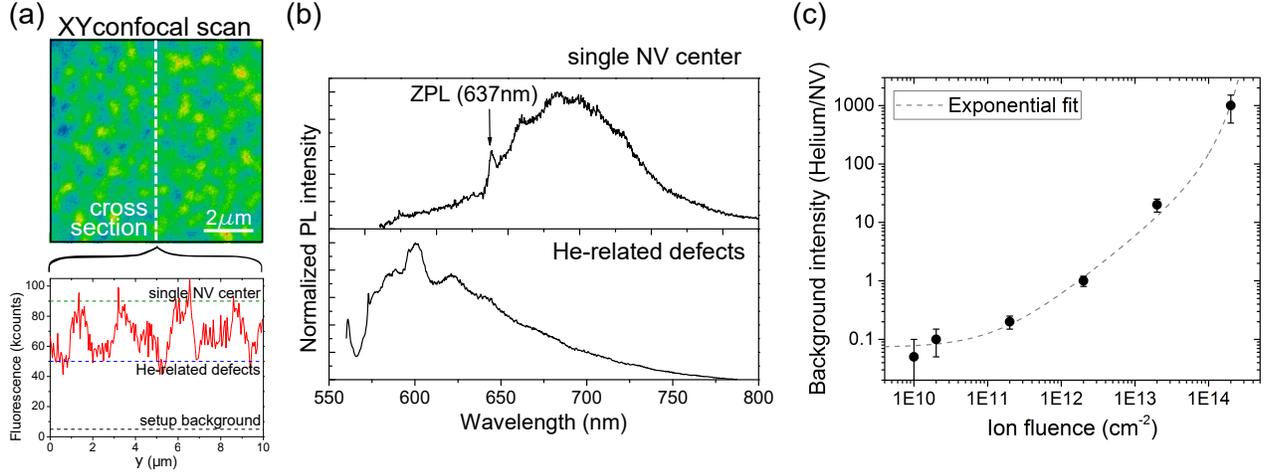}
\caption{\label{SI1} (a) An example of a confocal microscopy scan from the diamond surface irradiated with an ion fluence of $10^{12}\,$cm${}^{-2}$; a cross section (lower) reveils the helium-related background. (b) Spectra related to a single NV center (upper) and helium-related defects (lower). (c) Rate between the helium-related fluorescence and the count rate of a single NV center vs. the helium ion fluence.}
\end{figure}

A particular feature of helium irradiation was investigated using the described CVD diamond with low and homogeneous concentration of ingrown nitrogen impurities. In a low ion fluence regime - up to $10^{11}\,$cm${}^{-2}$ - only single NV centers were created, reflecting the low content of nitrogen in the used crystal. In contrast, by increasing ion fluences, a photoluminescence (PL) background was revealed concurrently with NV centers. As an example, the confocal microscopy PL scan from the diamond surface region irradiated with an ion fluence of $10^{12}\,$cm${}^{-2}$ is shown in figure \ref{SI1}(a). The related cross section from the XY scan (lower) unveils the increased PL background signal in comparison to the expected setup background. It is important to highlight that such helium-related PL signal is only present after thermal annealing. Such behavior is typical for a number of irradiation-related defects in diamond, which suggests a complex structure of these helium-related defects as well\cite{opticalpropDiamond}. The PL signal related to these defects was shown to increase rapidly with increasing ion fluence, setting a practical limit of approximately $10^{13}\,$cm${}^{-2}$ for resolving single NV centers by confocal microscopy technique, as shown in figure \ref{SI1}(c). In addition, the non-linear behavior of the helium-related background further supports the assumption that vacancy recombination mechanisms limit the number of vacancy available for the formation of NV centers with increasing ion fluence.   

\begin{figure}
\includegraphics{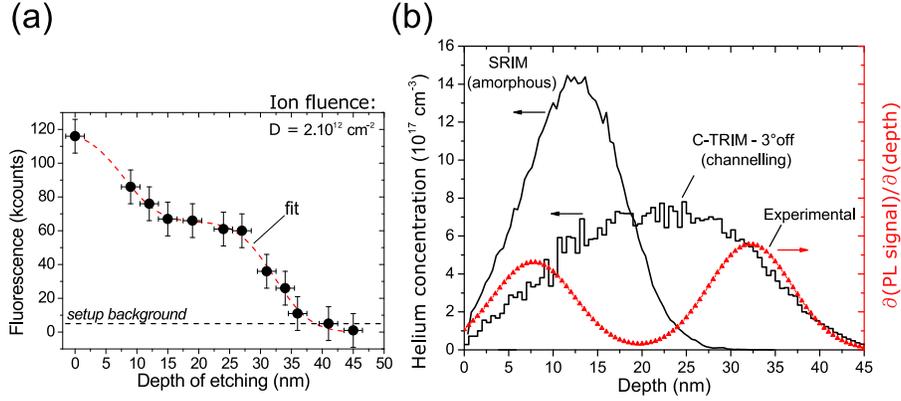}
\caption{\label{LHIF} (a) Experimental helium-related PL background signal vs. the etching depth. (b) The experimentally fitted profile is compared to corresponding SRIM and CTRIM simulations.}
\end{figure}

As a result of the described feature, profiling simultaneously the depth distributions of helium atoms and helium-induced NV centers becomes feasible. Each evaluation step comprised the removal of $1$ to $5\,$nm from the diamond surface by the mentioned plasma process followed by measurements of the helium-related PL background intensity and the areal density of NV centers by confocal microscopy technique, as presented in figures \ref{LHIF}(a) and \ref{figure4}(c) - main text, respectively. The experimental values were fitted for simplicity to Gaussian complementary error functions, as described in the main text.

In figure \ref{LHIF}(b), the experimentally fitted Gaussian distribution of helium-related defects is compared to correspondent atomic profiles generated by SRIM and CTRIM simulations, associated to ion implantation in an amorphous material and a crystal lattice, respectively. As it can be observed, the experimental profile shows a better agreement to the CTRIM profile with a $3^{\circ}$-off implantation angle, thus allowing \textit{ion channeling} to be resolved.

\section{Sensing Gd${}^{3+}$ by longitudinal spin relaxation}

Gadolinium (Gd${}^{3+}$) ions are known to be interesting spin systems, possessing a relatively high magnetic moment of S $= {}^{7}/{}_2$ as a result of seven unpaired electrons in the 4f orbital. Additionally, they exhibit a broad noise spectral density with significant intensity at the zero-field splitting frequency of NV centers\cite{SteinertGd,PelliccioneGdTip}. Since the spin-lattice relaxation time (T${}_1$) of NV centers is known to be sensitive to magnetic noise at frequencies comparable to its Larmor precession, external Gd${}^{3+}$ ions decrease the values of T${}_1$ from nearby NV centers.

For the described experiment, the layer containing Gd${}^{3+}$ ions was created by spin-coating a solution of Gadovist with a concentration of $1\,$M and drying it in air, yielding a homogeneous layer of approximately $1\,$\textmu m in thickness on the diamond surface. For the results presented in the main text, the following considerations were applied. Although the mean amplitude of random magnetic fluctuations is zero ($\langle {B} \rangle = 0$), statistical polarization would yield a non-zero root mean square field ($B_{RMS} = \sqrt{\langle {B}^2 \rangle}$). In the limit of a large number of ions, the magnetic field strength at the position ``$d_{NV}$'', the depth of an NV center, would be described as\cite{SteinertGd}:

\begin{equation}\label{BGd}
\langle {B}^2 \rangle = \frac{21 \times 10^3 \pi N_A C_{Gd}}{16 {d_{NV}}^3} {\left (\frac{{\mu}_0 {{\mu}_B}^2 g_{NV} g_{Gd}}{4 \pi \hbar} \right)}^2
\end{equation}

where $N_A$ is the Avogadro`s number, $C_{Gd}$ is the volumetric concentration of Gd${}^{3+}$ ions, $g_{NV}$ and $g_{Gd}$ are the electron g-factor of NV centers and Gd${}^{3+}$ ions, respectively. This can be related to the Gd${}^{3+}$-induced decrease in the spin relaxation time (T${}_1$) by\cite{SteinertGd}:

\begin{equation}\label{relax}
\frac{1}{{T}_{1,Gd^{3+}}} - \frac{1}{{T}_{1,int}} \approx \frac{2 f_t \langle {B}^2 \rangle }{{f_t}^2 + D^2}
\end{equation}

with ${T}_{1,Gd^{3+}}$ and ${T}_{1,int}$ being the spin-lattice relaxation times in the presence and absence of Gd${}^{3+}$ ions, respectively, $f_t$ being the fluctuation rate of the magnetic noise (GHz range)\cite{SteinertGd}, and D $= 2.870\,$GHz being the zero-field splitting frequency of NV centers. 

\begin{figure}
\includegraphics{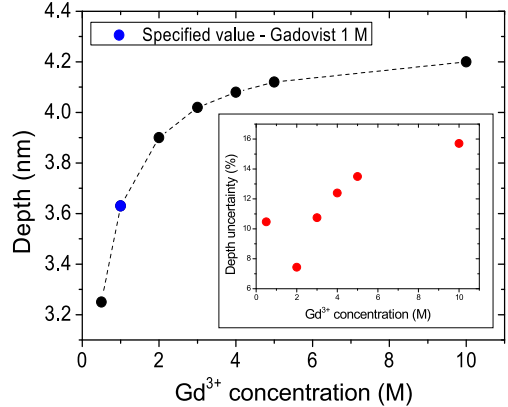}
\caption{\label{SI2} Estimation of the error in the depth of one individual NV center caused by the uncertainty in the concentration of Gd${}^{3+}$ ions in the surface layer inserted in the used model. The inset represents it as a percentage of the original value.}
\end{figure}   

As observed in figure \ref{SI2}, in the limit of high concentration of Gd${}^{3+}$ ions used in the experiment, an error of up to $10\,\%$ would be expected for uncertainties in ion concentration from $0.5$ up to $3\,$M. Therefore, the uncertainty in the calculated depths shown in figure \ref{figure3}(b) in the main text can be attributed mainly to variations in the T${}_1$ time measured before and after cleaning the sample from the Gd${}^{3+}$ layer. An example of the described relaxation measurements is presented in figure \ref{figure3}(a) from the main text. The results show a difference in T${}_1$ times corresponding to the initial and clean surface conditions. In addition, the corresponding T${}_2$ time was shown in the inset of figure \ref{figure3}(a). The T${}_2$ time did not change significantly in the presence of Gd${}^{3+}$, which is expected since the intrinsic noise related to the transverse spin relaxation (kHz to MHz range) is usually high due to the interaction with the diamond surface\cite{RosskopfSpinSurf,SteinertGd}.

\begin{figure}
\includegraphics{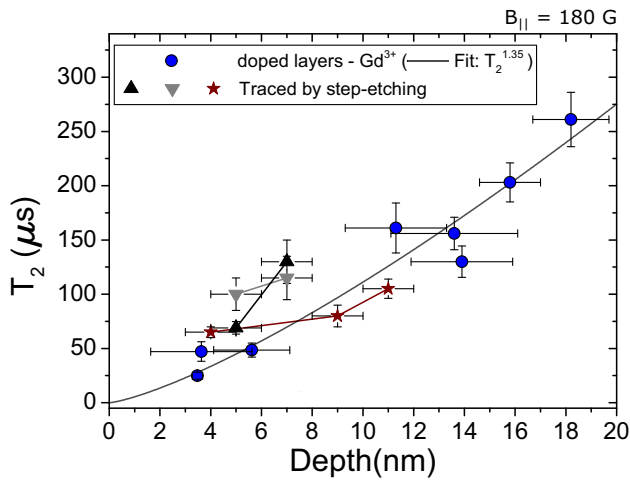}
\caption{\label{SI3} T${}_2$ time vs. depth of individual NV centers; in addition to depth estimated by sensing Gd${}^{3+}$ ions at the diamond surface as described in the main text (circles), individual NV centers were traced through the etching steps until their disappearance (triangles and stars).}
\end{figure}

T${}_1$ times were fitted by a single exponential in the form $A{e}^{{}^{\tau}/{}_{T_1}}$\cite{JarmolaT1}. In order to consider the electron spin echo envelope modulation (ESEEM) due to the presence of ${}^{13}$C spins in the diamond lattice, T${}_2$ times were fitted using the approach described by B. K. Ofori-Okai \textit{et al.}\cite{Ofori-OkaiSPVeryShallow}.

To evaluate possible effects of the oxygen soft plasma process, individual NV centers were traced through the calibration etching steps until their disappearance. At this point, their relative depth could be estimated by the amount etched in each step of the sequence. The corresponding T${}_2$ times were measured and are shown in figure \ref{SI3} by triangles and stars, representing three different NV centers. Importantly, the depth values estimated by the step-etching are in good agreement with the values calculated by sensing Gd${}^{3+}$ ions. This reinforces that the spin and optical properties of near-surface NV centers are not affected by the plasma process.

\begin{figure}
\includegraphics{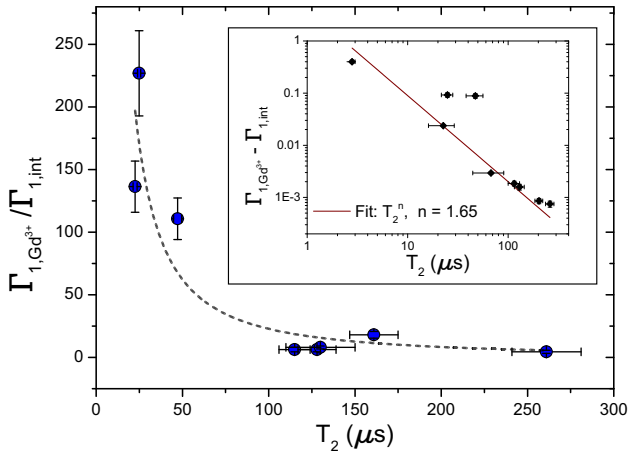}
\caption{\label{SI4} Rate between the Gd${}^{3+}$-induced and intrinsic spin relaxation vs. corresponding T${}_2$ times; inset shows the difference between the mentioned relaxations used to estimate the depths of individual NV centers.}
\end{figure}

To illustrate the consistency of the used spin relaxation model, the rate between the Gd${}^{3+}$-induced and the intrinsic spin-lattice relaxation in relation to T${}_2$ times of individual NV centers is shown in figure \ref{SI4}. The inset shows the difference in the mentioned relaxations ($\Gamma _{1,Gd^{3+}}$ - $\Gamma _{1,int}$) used to estimate the depth of individual NV centers, as presented in the main text. The fitting curve in figure \ref{SI4} depicts a good agreement between the experimental results of coherence times and the expected behavior by the used model (T${}_2^{1.35}$ fit shown in figure \ref{figure3}(b) in the main text).

\end{suppinfo}

\bibliography{F.F.Oliveira_IR}

\end{document}